\documentclass[prb,twocolumn,showpacs]{revtex4}
\usepackage{graphicx}
\usepackage{pstricks}

\begin{document}
\title{
Curvature Effects on Optical Response 
of Si nanocrystals in SiO$_2$ having Interface Silicon Suboxides 
}

\author{\firstname{Pierre} \surname{Carrier}} 
\affiliation{Minnesota Supercomputing Institute and Department of Computer Science 
\& Engineering, University of Minnesota, Minneapolis, MN 55455}
\email{carrier@cs.umn.edu}

\date{\today}
 
\begin{abstract}
Several models of oxygenated and hydrogenated surfaces of Si
quantum wells and Si nanocrystals of 
variable shapes have 
been constructed in order to assess curvature effects on energy gaps 
due to the three Si-suboxides.
Si-suboxides in partially oxydized models of 
nanocrystals of spherical shapes (or quantum dots) are shown 
to reduce energy gaps compared to the hydrogenated nanocrystals, 
consistent with previous results in the literature.
This trend is shown to be reversed when planar interfaces are formed in Si-NC 
inside SiO$_2$ thin films, as in Si quantum wells.
At planar interfaces (or surfaces) the electronic charge density is shown to become 
extended and to distribute
among the Si-suboxides,
thus generating along these planar interfaces extended, or delocalized,
states. 
This delocalization of the electronic states then
increase the 
energy gap compared to 
equivalent hydrogenated interfaces (or surfaces).
Determination of geometric effects of curvature on the band 
gaps are based on density functional theory (DFT)
using the real-space numerical approach implemented in the 
``Pseudopotential Algorithm for Real-Space Electronic
Structure'' (PARSEC) program.
A method for measuring interface effects due to Si-suboxides in Si-NC 
in SiO$_2$
is also suggested: by comparing photoluminescence (PL) between as-grown and 
annealed Si-NC samples.
The DFT calculations suggest that blueshift of more than 0.2 eV of the 
PL should be observed in as-grown 
samples having Si-suboxides at their planar interfaces, in 
comparison to annealed samples above 950$^{\circ}$C
that have fewer or no Si-suboxides once annealed, providing that the bulk of Si-NC remains unaltered.
\end{abstract}

\maketitle

\section{Introduction}

Silicon/SiO$_2$ quantum wells (Si-QW) \cite{Lu} as well as silicon 
nanocrystals in 
SiO$_2$ (Si-NC)\cite{Thogersen,Heitmann} are examples 
of \emph{all-silicon}\cite{Cho,Svrcek} 
optoelectronic materials applicable to future generation high-efficiency
photovoltaic multi-junctions tandem solar cells.\cite{HandbookPV}
Such devices are non-toxic and made of the two most abundant 
elements: silicon and oxygen. 
Both Si-QW and Si-NC structures constitute
essentially artificial
metastable forms of bulk-SiO$_2$. 
However, quantum confinement of
silicon in the form of Si-QW or Si-NC 
drastically modifies the electronic and optical properties of bulk-Si, 
and consequently, bulk-SiO$_2$ as well.
Quantum confinement\cite{Ren}
transforms the indirect nature of the
band gap of bulk-silicon \cite{Madelung} into a direct band gap 
material\cite{Carrier}
when the size of confinement of silicon is of the order of the wavelength 
of its
electronic wavefunction (i.e., few
nanometers), thus enhancing the photoluminescence (PL) of Si.
Enhancements of PL due to quantum confinement in silicon were originally 
reported in 
porous-Si,\cite{Canham} in Si-QW,\cite{Lu} and in Si-NC 
.\cite{Linnros}
Another well known characteristic of quantum confinement is that the 
excitation energy increases with increased confinement. 
For instance, 
Si-QW are confined in one direction (1D-confinement) and the difference between 
the conduction band minimum (CBM) 
and valence band maximum (VBM), i.e., the band gap of the crystal, can vary 
from $\sim$1.5 eV for very thin wells (containing 2-3 monolayers of Si)  
down to  $\sim$1.1 eV for larger wells. 
Si-NC, on the other hand, are further confined in the three directions 
(3D-confinement) and the difference between the 
lowest unoccupied molecular orbital (LUMO) and the
highest occupied molecular orbital (HOMO) 
i.e., the HOMO-LUMO gap of the cluster, can vary from values just below the 
band gap of SiO$_2$ (9 eV), when few silicon atoms form large enough
clusters inside SiO$_2$,  down to the band gap of bulk-Si 
(1.1 eV).
In practice, the reported HOMO-LUMO gaps in Si-NC are generally less 
than $\sim$4 eV. 

The Si/SiO$_2$ interface between the semiconductor Si and the dielectric 
medium SiO$_2$ in Si-QW or Si-NC is  the practical instrument for generating 
quantum confinement in physical device.
In other words, the
Si/SiO$_2$ interface is inherent to the physics of Si quantum confinement and
the influence of the atomic structure  at the Si/SiO$_2$ 
interface cannot be neglected. 
Determining the exact interface structure 
between Si and SiO$_2$ is however a difficult task,  mainly because SiO$_2$ 
is amorphous. 
It is in part
due to the large lattice mismatch between the two bulk materials:
e.g., the lattice parameter of crystalline Si is  $a$=5.435 \AA\ while
that of SiO$_2$ in 
the  ideal $\beta$-crystobalite phase is a=7.16 \AA.\cite{CarrierDBMBOM}
For this reason SiO$_2$ grown on crystalline-Si wafers is always amorphous.
The planar SiO$_2$/Si(100) interface structure has extensively been probed 
in the past especially for purposes of characterization of the metal-oxide 
semiconductor field-effect transistors (MOSFET) structure.
In general, an SiO$_2$/Si interface can be probed using  
high energy \textit{photons} 
(e.g., $\sim$100 eV) that excite Si$_{2p}$ core \textit{electrons}, i.e., using 
X-ray \textit{photoelectron} spectroscopy (XPS).\cite{Fuggle}
XPS gives in general information on the chemical states between
two different chemical environments (Si and SiO$_2$ in this case) using 
binding energy and Auger electronic kinetic energy 
as parameters.\cite{Fuggle, Mejias, Thogersen}
The shift of the XPS spectrum in 
silicon-on-insulator (SOI) shows that planar SiO$_2$/Si interfaces always 
contains three Si-suboxides, which are by definition three silicon 
atoms bonded to 1, 2, and 3 oxygens, respectively referred to as Si$^{1+}$, 
Si$^{2+}$, and Si$^{3+}$.
It has recently been shown, using XPS experiments as in the SOI, 
that the three 
Si-suboxides are also at the Si/SiO$_2$ interfaces of Si-QW\cite{LuTay} 
and Si-NC.\cite{Thogersen}
%FIGURE_1
\begin{figure}[tb]
   \centering
   \includegraphics[width=3.0in]{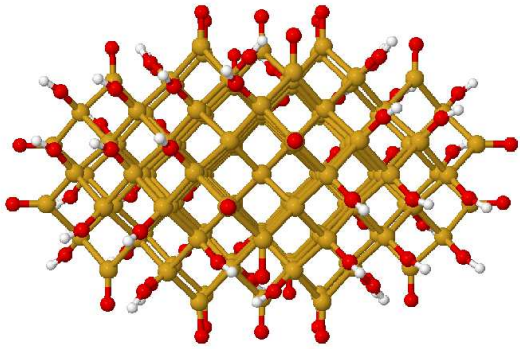}
   \caption{(Color online) Side view of the disk-like Si-QD containing 
126 Si atoms. The surface is made of 26 double-bond oxygens and 48 hydroxides.
Silicon atoms are light (yellow) spheres, oxygen atoms are dark (red) spheres,
and hydrogen atoms are small (white) spheres.}
   \label{Fig1}
\end{figure}

Theoretical investigations that describe effects of oxygen on the 
band gap or the HOMO-LUMO gap in Si confined structures are more or 
less divided into two main 
groups: one group of investigations focusses on the structure of the 
\emph{planar} Si/SiO$_2$ 
interface, of particular interest in MOSFET or SOI, and in the Si/SiO$_2$ 
QW (see Ref.\ \onlinecite{Carrier} and references therein, or the 
Review on Si/SiO$_2$ superlattices in
Ref.\ \onlinecite{Zheng}).
A second group of investigations focuses on oxygen passivation 
at the surface of \emph{spherical} Si quantum dots 
(QD),\cite{Wolkin, Vasiliev, Puzder, Nishida, Koenig} of 
particular interest for Si-NC in SiO$_2$ thin films. 
This somewhat arbitrary division may be attributed to the difference in the 
treatment of boundary conditions (BC) in numerical simulations. 
Quantum wells 
are periodic systems, which is an ideal setting for Fourier basis 
(i.e., a plane wave set) used in general for simulation of crystals. 
Quantum dots are confined systems (with zero BC) which 
is an ideal setting for finite basis (e.g., gaussian set) used in 
general for simulations of molecules. 
Most of the results on the role of oxygen in Si
confinement are thus usually confined to their respective 
geometries and rarely geometrical effects are juxtaposed.
This article is intended to bridge this gap.
It also resolves an apparent contradiction 
concerning the role of oxygen on band gaps for various
interface topologies (planar or spherical),  modeled using various BC.

The apparent contradiction on the role of oxygen in Si confined structures 
is the following:
Surfaces of spherical Si-QD passivated with oxygen are reported to 
always \emph{reduce} 
the HOMO-LUMO gap in comparison to equivalent hydrogen passivated 
surfaces.\cite{Puzder, Vasiliev, Wolkin}
This reduction of the energy gap is usually attributed to a 
larger localization of states at the oxygen sites of the 
surface.\cite{Puzder}
On the other hand,
earlier results published by the author on Si-QW models containing the three 
Si-suboxides at the surface (or interface) of the well were shown to 
\emph{increase} the band gap compared to equivalent hydrogenated 
QW models.\cite{Carrier}
This article first confirms the previous 
result \cite{Carrier}  
using a different numerical framework described below.
It also shows that the energy gap in Si-NC does not always decrease due to oxygen and that 
this gap variation depends also
on the shape of the nanocrystal, especially
if the nanocrystal has planar interfaces over just few nanometers, as shown below.
Finally, a description of the mechanism 
leading to different 
variations of band gaps in Si-QW and Si-QD due to Si-suboxides is suggested. 
The mechanism is the following: 
\emph{Collective interactions between neighboring Si-suboxides is 
obviously increased
at a planar surface compared to a large curvature nanocrystal.
This increased interaction leads to extended,
delocalized, electronic states at the surface,
thus implying an \emph{increase} of the band gap compared to 
similar hydrogenated surfaces.} 
The interaction of neighboring Si-suboxides is less intense as the size of the 
Si-QD diminishes, due to simple curvature effects, which explains in part why 
oxygen atoms in Si-QD have reduced energy gaps compared to
bare hydrogenated Si-QD (see Fig.\ \ref{Fig3} and discussion below).
This collective effect that generates extended, or delocalized, electronic 
states at the interface or surface
(see Fig.\ \ref{Fig4} and the discussion below)
implies in particular that studying an individual 
Si-O atomic configuration or bond (e.g., a particular double-bond oxygen or
bridge oxygen, or any particular Si-suboxide configuration, alone) in 
Si confined 
structures is indeed essential, but is not sufficient for explaining the 
whole process of 
energy gap variations due to surface 
or interface oxygen atoms. 
%FIGURE_2
\begin{figure}[tb]
   \centering
   \includegraphics[width=2.1in]{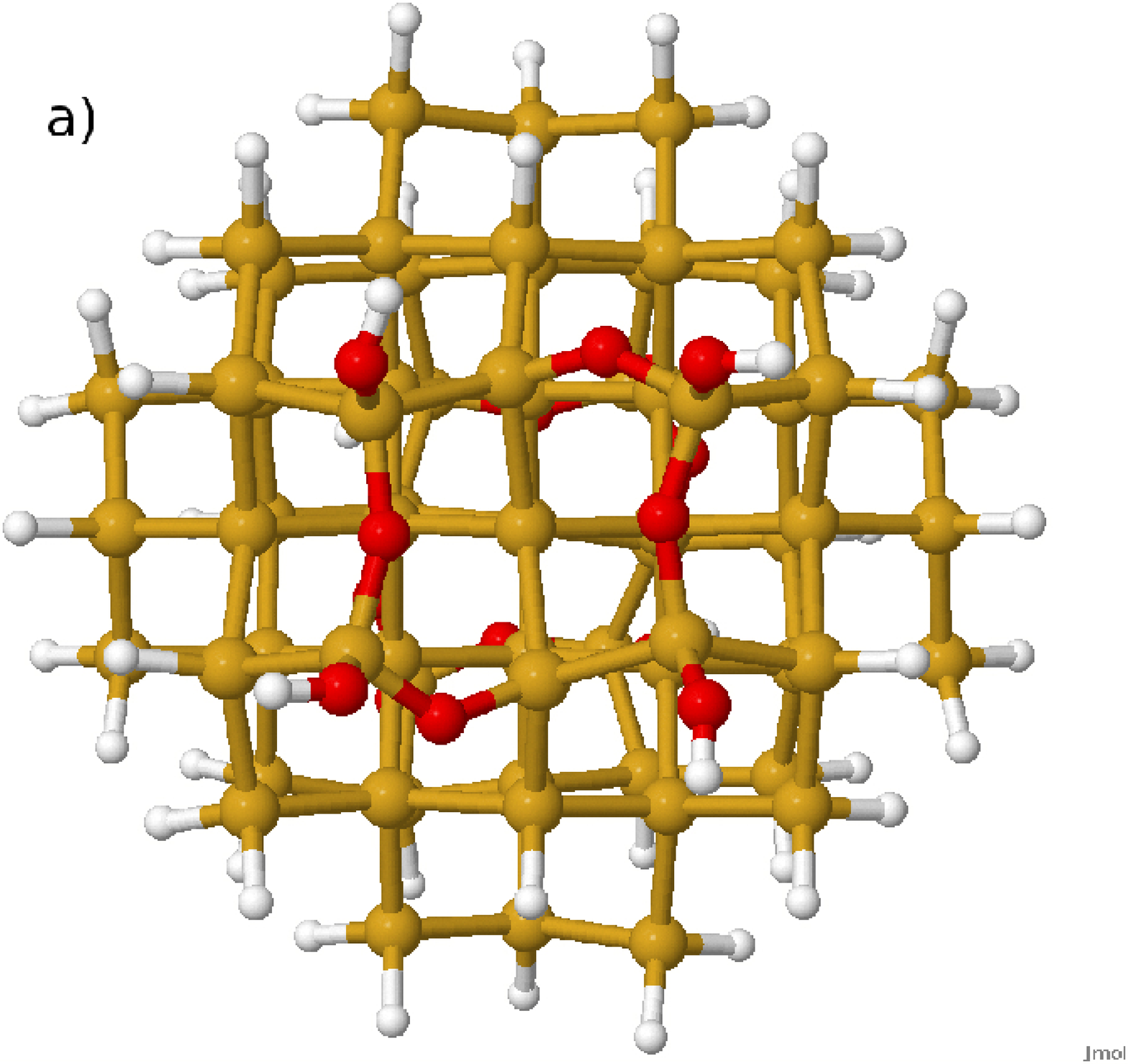}
   \includegraphics[width=2.1in]{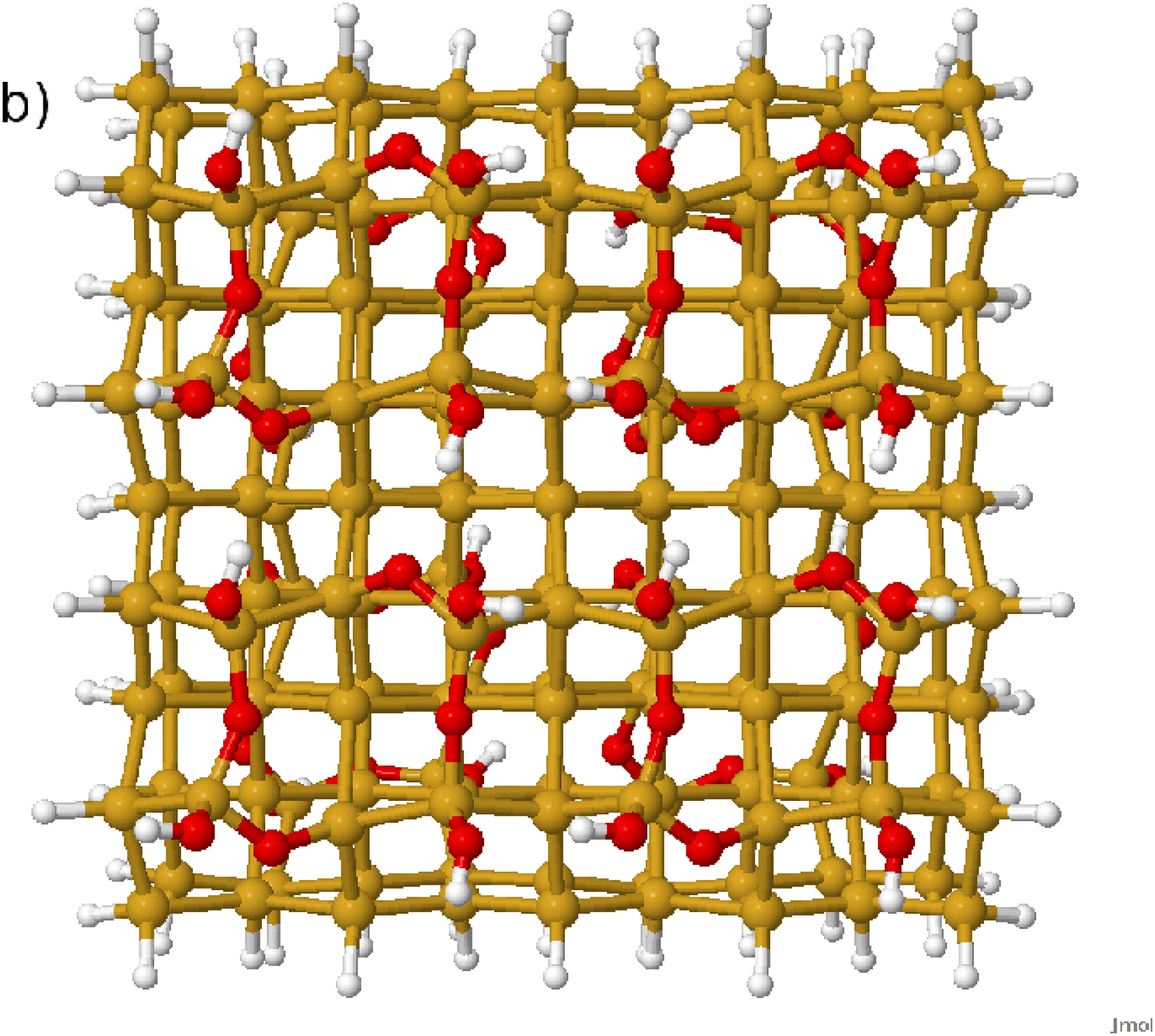}
   \includegraphics[width=2.1in]{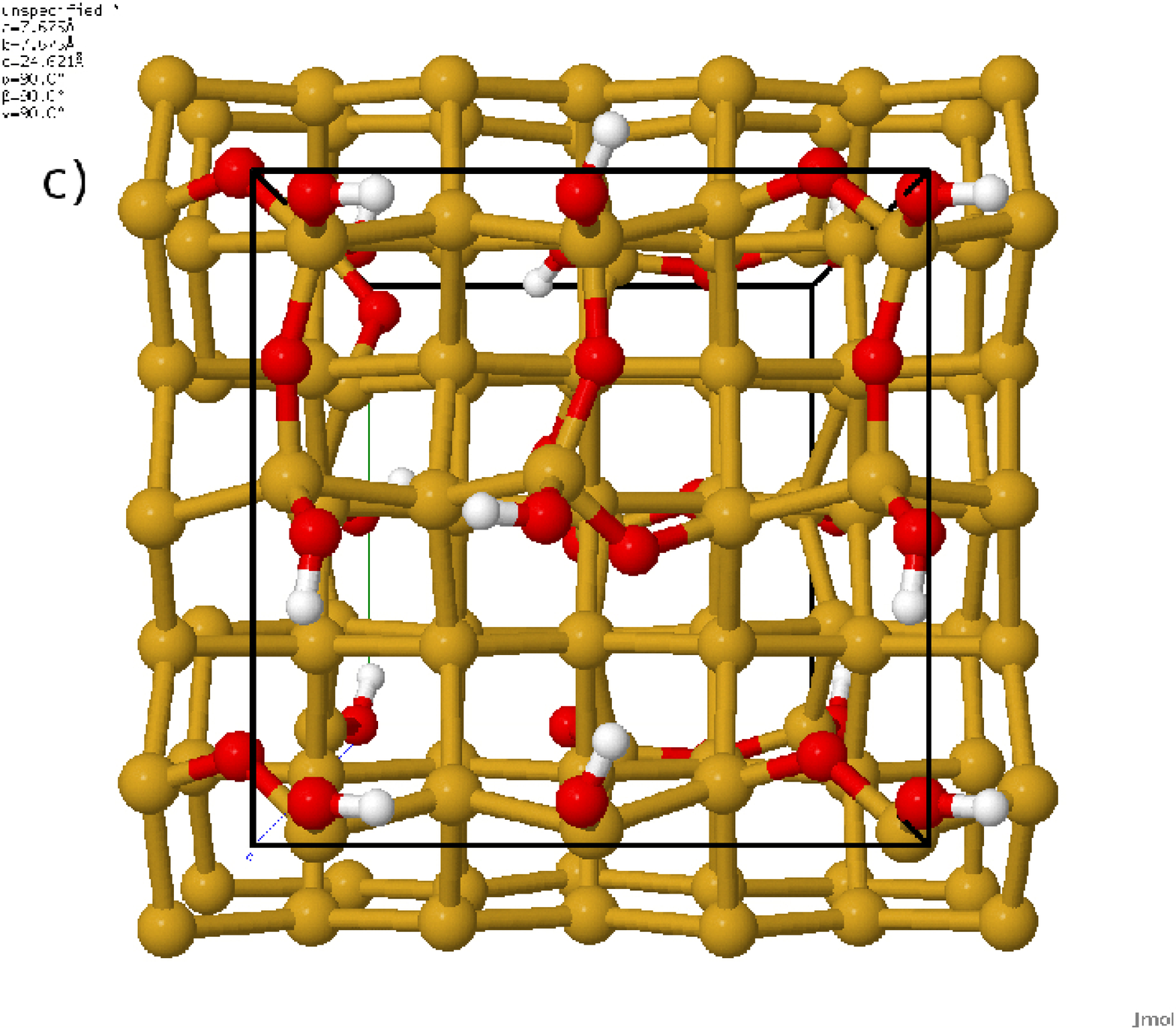}
   \caption{(Color online) 
Top view of three models containing the 3 Si-suboxides, with variable 
geometries.
(a) The Si-QD model is made of 66 Si atoms, containing 4 of each 
Si$^{3+}$, Si$^{2+}$, and Si$^{1+}$ suboxides, for a total of 16 O atoms, 
and 56 H atoms (among which 8 are bonded to O atoms).
(b) The planar Si(100) nanocrystal 
model is made of 178 Si atoms, 16 of each 
Si$^{3+}$, Si$^{2+}$, and Si$^{1+}$ suboxides, 
for a total of 64 O atoms, and 104 H atoms.
(c) The Si-QW model is made of 52 Si atoms and the same number of O and H 
atoms as in the Si-QD.
Image mirror atoms outside the periodic unit cell have been repeated for 
clarity.
All three models have 9 Si monolayers that separates the Si-suboxides visible
at the top 
to the invisible Si-suboxides at the bottom.
Colors are defined in Fig.\ \ref{Fig1}.
Interface model is based on the surface model from 
Pasquarello \textit{et al}.\cite{Pasquarello}
}
   \label{Fig2}
\end{figure}

\section{First-principles method}

The Kohn-Sham equation\cite{KohnSham}
is resolved 
with the ``Pseudopotential Algorithm for Real-Space Electronic Calculations'' 
(PARSEC) program.\cite{PARSEC1, PARSEC2, PARSEC3}
Using real-space instead of Fourier-space  algorithms for solving the Kohn-Sham
equation facilitates treatments and modifications of BC. 
It also facilitates comparisons of eigenenergies 
between different BC 
(periodic BC for the Si-QW and zero BC for the Si-QD)
using \emph{identical} numerical parameters.
All pseudopotentials (PP) are constructed using the Troullier-Martins 
algorithm.\cite{TroullierMartins}
Silicon, oxygen, and hydrogen PP are 
constructed using 4, 6, and 1 valence electrons and
the PP radii cutoffs are 2.782 a.u. (using local $p$ state), 
1.452 a.u. (using local $s$ state), and 1.30 a.u. (using local $s$ state), 
respectively.
Atomic relaxations are performed using the Broyden-Fletcher-Goldfarb-Shanno 
(BFGS) method.\cite{Nocedal}
Forces are less than 0.01Ry/a.u.\ ($\sim$0.001 eV/\AA) corresponding to 
energy gap errors less than $\sim$0.005 eV, tested on the model shown in 
Fig.\ \ref{Fig1}. 
The real-space mesh is set to 0.45 a.u.\ for both periodic and zero 
BC which gives an error less than 0.00001 eV on the total energy per atom.
The smallest 
supercell for the quantum wells  has dimensions 
7.675\AA $\times$ 7.675\AA $\times$ 
24.621\AA, which corresponds to a relatively small Brillouin zone (BZ).
Tests on the model depicted in Fig.\ \ref{Fig2}(c) using 4$\times$4$\times$2 
\textbf{k}-points and using
2$\times$2$\times$1 \textbf{k}-points for the BZ sampling show that the 
latter constitutes a sufficiently fine $\textbf{k}$-point mesh.
All structural relaxations and band gap evaluations have 
thus been performed using  2$\times$2$\times$1 \textbf{k}-points in the BZ.
With these numerical parameters the size of the real-space Hamiltonian $\cal H$ 
contains $\sim$2 millions 
elements in the largest model [Fig.\ \ref{Fig2}(b)] and corresponds to 
finding 600 occupied eigenstates 
(2 electrons per levels, no spin-orbit interactions).
Diagonalization of $\cal H$ and Chebyshev filtering\cite{Saad} are performed 
in parallel using Message Passing Interfaces (MPI) between 
256 processors on an 
\textit{SGI Altix XE 1300 Linux Cluster} or 128 processors on 
an \textit{IBM Bladecenter Linux cluster}. 
For example, the largest $\cal H$
distributed among 128 processors requires less than 400 Mb 
of memory per processors, a clear advantage of the real-space approach over 
other methods.

The exchange-correlation functional used throughout this work is the local 
density approximation (LDA).\cite{KohnSham} 
The LDA is chosen for purposes of comparison 
with previous works.\cite{Carrier} 
It is also chosen because the LDA band 
gap error in bulk-Si, caused by the self-interaction 
potentials, is well known and amounts to approximately 
-0.60 eV using the pseudopotential approach (the correction being slightly 
larger 
when using \emph{all-electrons} methods\cite{CarrierDBMBOM}).
Moreover, the LDA band gap error 
seems to have little impacts on the band gap variations
due to 
confinement in previous Si/SiO$_2$ QW models, where good agreement with 
experiments on
band gap variations were reported.\cite{MRS_Lockwood} 
Nevertheless, rigidly 
shifting the energy gaps corresponds to assuming that the LDA 
energy gap error is independent of the chemical species in use. 
Such 
assumption might underestimates energy gap differences between SiOH and SiH  
models,  based on the observation that the LDA band gap error 
of SiO$_2$ in the $\beta$-cristobalite phase is much larger than in 
silicon.\cite{CarrierAppSurf} 
Therefore, the reported LDA  blueshifts, below, constitute lower bounds of the 
full effects. 
%FIGURE_3
\begin{figure}[tb]
   \centering
   \includegraphics[width=3.0in]{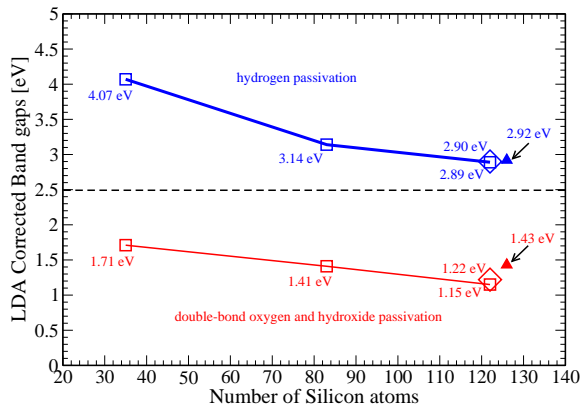}
   \caption{(Color online)
LDA HOMO-LUMO gap results for the first set of models. 
(To retrieve calculated LDA energy gap value 
subtract 0.6 eV ---the LDA band gap correction for Si--- to the numbers 
in the graph.)
The horizontal dotted line at 2.5 eV arbitrarily separates two general 
domains, one containing gaps of hydrogenated models (above the 
dotted line) and one containing gaps of the double-bond oxygen and 
hydroxide models (below that line).
Square symbols are connected in order to indicate the trends of 
energy gaps in both hydrogenated (thick blue line) and oxygenated 
(thin red line) spherical Si-QD, as discussed in the text. 
Diamond and filed-triangle 
symbols are energy gaps of oblong and disk-like Si-QD with hydrogen 
or Si-suboxides,
respectively. 
The difference of 4 Si atoms between the disk-like (126 Si atoms) dots and 
oblong or spherical dots (122 Si atoms) is kept so as to 
ensure uniformity of shapes.
}
   \label{Fig3}
\end{figure}

\section{Si Confinement Models}

Two sets of models have been constructed in this work and are 
fully described in the next
two paragraphs.
 
The first set is 
used for estimating general trends related to effects of shapes on the energy gaps 
using spherical, oblong, and disk-like Si-QD models. 
In these models the 
entire surface is uniformly passivated with hydrogen, or passivated with 
double-bond oxygen atoms and hydroxides.  
Three sizes of hydrogenated spherical Si-QD models have been constructed 
for evaluating general energy gap trends, with the number of 
[silicon, hydrogen] atoms being respectively [35, 36], 
[83, 72], and [122, 100] in each of the 3 sizes of models.
Three other oxygenated spherical Si-QD models with [35, 30, 24], 
[83, 51, 30], and [122, 64, 28] 
[silicon, oxygen, hydrogen] atoms have also been constructed for comparison.
They contain 6, 21, and 36 double-bond oxygens and 24, 30, and 28 
hydroxides, respectively.
The diameter (i.e., the largest distance between any two silicon atoms) of these 
dots are respectively about 1.09 nm, 
1.38 nm, and  1.54 nm.
Comparisons of shapes have been performed only on the largest dots.
The hydrogenated oblong dot contains [122, 106] [silicon, hydrogen] atoms,
a size directly comparable to the largest
spherical dot.
The oblong model has a shortest diameter 
of 1.09 nm 
which is  also
comparable to the diameter of the smallest spherical dots, 
and has a largest diameter of 2.46 nm.
The oxygenated oblong dot contains [122, 68, 30] [silicon, oxygen, hydrogen] atoms. 
It corresponds to only 
2 additional double-bond oxygens and 2 additional hydroxides than same size 
spherical dot described above.
Finally, the hydrogenated disk-like dot model contains 
[126, 100] [silicon, hydrogen] atoms and
its oxygenated disk-like dot counterpart contains [126, 74, 48] 
[silicon, oxygen, hydrogen] atoms.
It corresponds to having 10 less double-bond oxygen atoms than 
its spherical counterpart and 
20 more hydroxides, which are actually significant changes of bond types
compared to all
the other models of similar size.
Nevertheless, this series of models is intended to provide general trends due to
a uniform geometry of the structure.
The diameter of the disk is about 1.92 nm and its thickness at the center is 
about 1.09 nm, which makes it again 
comparable to the smallest spherical dot containing 35 Si atoms.
Each model is constructed so as to ensure uniformity of the structure.
Fig.\ \ref{Fig1} depicts the oxygenated disk-like models.
Results are further discussed in the next Section along with Fig.\ \ref{Fig3}.
 
The second set of models are partially oxydized nanostructures containing
the three 
Si-suboxides at the surface, as in XPS, and are derived from the superlattice
models in Ref.\ \onlinecite{Carrier}, depicted in Fig.\ \ref{Fig2}(c).
The purpose of this second set of models is to complement the first set.
The first set was
uniformly passivated but contained unrealistic oxygen bonds, while the second
is oxydized only partially but contains realistic oxygen bonds
and is intended to describe precise 
trends caused by a given structural configurations of 
Si-suboxides after modifying only the geometries.
All oxygenated models are depicted in Fig.\ \ref{Fig2} and energy gaps 
are tabulated in Table \ref{Table1} as further described in the next Section.
Fig.\ \ref{Fig2}(a) shows the spherical quantum dot.
The diameter of the dot is 1.14 nm.
It contains [66, 16, 56] [silicon, oxygen, hydrogen] atoms, 
with 4 of each of the Si$^{1+}$, Si$^{2+}$, and Si$^{3+}$ suboxides.
A hydrogenated spherical Si-QD models is constructed for comparisons of energy
gaps (not depicted) and contains [66, 64] [silicon, hydrogen] atoms.
Fig.\ \ref{Fig2}(b) now shows the Si-NC that has two oxygenated planar 
surfaces of dimension 1.56 nm $\times$ 1.56 nm 
separated by 1.14 nm, or 9 monolayers of Si.
Notice that the thickness of the planar nanocrystal is identical to the 
diameter of the 
quantum dot of Fig.\ \ref{Fig2}(a).
It contains [178, 64, 104] [silicon, oxygen, hydrogen] atoms, with 16 of each 
of the Si$^{1+}$, Si$^{2+}$, and Si$^{3+}$ suboxides.
The two models in Fig.\ \ref{Fig2}(a) and (b) were
constructed starting from the periodic system depicted in 
Fig.\ \ref{Fig2}(c),
one by generating a sphere directly from the superlattice model, and the other
by first doubling the cell in both $x$--$y$ directions of the periodic supercell 
and then passivating
all dangling bonds with hydrogen on the sides of the Si-well.
All models in Fig.\ \ref{Fig2} have therefore identical structural atomic
configurations for the Si-suboxides, and differ only by their 
geometries: (a) is a spherical quantum dot,
(b) is a planar nanocrystal, and (c) is a crystalline superlattice.
All atomic positions have been structurally relaxed in their respective
geometries, using zero BC for the models in Fig.\ \ref{Fig2}(a) and (b) and using
periodic boundary conditions for the model in Fig.\ \ref{Fig2}(c).
Although bond lengths and bond angles surrounding the Si-suboxides
vary slightly between geometries, those variations 
do not have a significant effect on the energy gaps reported below.
For example, 
as stated above, the Si-QD model in Fig.\ \ref{Fig2}(a)
was constructed starting from the atomic
positions of the Si-QW in Fig.\ \ref{Fig2}(c) and 
such structural relaxation from the atomic
positions of the QW suboxides towards these of 
the QD suboxides modify the LDA band gap by only 
0.01 eV ---from 2.35 eV to 2.36 eV--- for a
total energy variation per atom of 0.25 eV after the full structural relaxation is completed.
It indicates that the energy gap variations inside the different geometries are independent of
specific atomic structural relaxations. 
That is, the variations of the energy gaps reported in Table \ref{Table1}
are global topological effects and are not due to particular structural atomic  configurations.
 
Notice that two general sets of models needed to be designed in this study for the reason that
a Si-QD or 
Si-NC model containing the
three Si-suboxides, as in the second set, and being
 uniformly
distributed on an entire curved surface, as in the first set,
has yet to be developed: such model has not been reported in 
the literature essentially because of its high complexity.

All atomic positions of the structurally relaxed models presented in this 
article are available from the author, upon request.
\begin{table}[tb]
\caption{ Band gaps and HOMO-LUMO gaps (corrected from the LDA error) 
of the three Si-suboxide models
depicted in Fig.\ \ref{Fig2}, compared to the energy gaps of their 
corresponding hydrogenated models.
Subtract 0.6 eV to retrieve the LDA values. Notice that hydrogenated 
Si(100)2X1 and 1X1 QW surfaces 
(last two lines) have 
similar band gaps, and that both have lower band gaps than that of the 
Si-suboxides QW. \label{Table1}
}
\begin{ruledtabular}
\begin{tabular}{lrrr}
Surface passivation & Si-H & Si-suboxides & difference \\ \hline
Si-QD                      & 3.17      &  2.96     &  -0.21 \\
planar Si(100)-NC   & 2.42      &  2.46     &  +0.04 \\
Si(100) 2X1-QW      & 1.53      &  1.82     &  +0.29 \\
Si(100) 1X1-QW      & 1.60      &    -      &  +0.22\\
\end{tabular}
\end{ruledtabular}
\end{table}

\section{Results}
 
Figure \ref{Fig3} shows the \emph{corrected} 
LDA HOMO-LUMO gap results for the first set of models. 
Five observations stand out on this graph.
(1) The quantum confinement effect is clearly noticeable: the energy gaps 
increase with increased confinements in both hydrogenated and oxygenated dots.
(2) The presence of double-bond oxygens dramatically reduce the energy gaps 
compared to that of hydrogenated Si-QD of any shape. 
This was already noticed in 
previous works using spherical dots.\cite{Wolkin, Puzder}
This significant drop of the energy gap due to double-bond oxygens
is also observed in Si-QW. 
For instance, 
an Si-QW containing 9 monolayers (ML) of Si passivated with hydrogen has 
an LDA band gap of 1.00 eV (1.60 eV, with LDA correction) while the same 
well passivated with a double-bond oxygen gives an LDA band gap of 
0.36 eV (0.96 eV, with LDA correction, i.e., below bulk-Si indirect band gap!).
Although never observed to exist in silicates, the double-bond oxygen is 
however a useful surface and interface model for 
determining general trends on the variations of band gaps using simple uniform models
chosen to have charge neutrality.
(3) The energy gap variations with increased confinement of oxygenated  
spherical dots (red thin line in Fig.\ \ref{Fig3}) is severely damped compared to that of 
hydrogenated 
dots (thick blue line). 
For instance, energy gaps of hydrogenated dots increase  by 1.18 eV when dot 
size decreases from 122 to 35 Si atoms, while energy gaps of oxygenated dots 
increases by about half this value, 0.56 eV, for equivalent decrease in dot size.
Smaller dots have larger curvature; larger 
curvature implies larger spacing between neighboring oxygens, which reduces 
interaction among neighboring oxygen atoms and consequently
may increase localization of the electonic states.
This suggests that a reduction of 
interaction between neighboring oxygen atoms at the surface further reduces
the energy gap in comparison to hydrogenated Si-QD, 
assuming that increasing the size of dots from 1.09 nm to 1.54 nm does not
significantly affect gap reduction due to size effects, i.e., the ratio of surface to bulk Si atoms
[see also item (5) below where size effects are excluded].
(4) The shape of \emph{hydrogenated} Si-QD does not radically affect the 
energy gaps. 
For instance, comparison in Fig.\ \ref{Fig3} of energy gaps for the 
hydrogenated spherical (square symbol), oblong (diamond symbol), and 
disk-like (filled-triangle symbol) Si-QD of 
similar size (122 or 126 silicon atoms) show that the trends along the thick blue line
remain clearly unbroken.
Moreover, 
the smallest diameter of the oblong and disk-like dots corresponds approximately
to the diameter of 
the smallest spherical dot ($\sim$1.09 nm), as described above, while energy gaps
of the disk-like and the smallest spherical dots
differ drastically (by 1.15 eV).
This suggests that the smallest diameter of non-spherical
quantum dots, which could have been viewed as the important 
confinement parameter for a given surface
passivation, does not actually define the energy gap.
It is rather the number of Si
in the whole dot that defines the gap.
Note that Ref.\ \onlinecite{Justo} describes a similar trend concerning 
surface shapes in 
quantum wires.
On the other hand, \emph{oxygenated} Si-QD of different shapes show a slight
variation of energy gaps with shapes, i.e.,
the trend along the thin red line is broken.
These general trends indicate that shapes of Si-QD must be taken 
into account especially 
when considering oxygen passivation of surface and interface.
(5) More precisely, disk-like oxygenated Si-QD  show a noticeable 
\emph{increase} of the band gap, by $\sim$0.2 eV above that of the other two
models of similar sizes but different shapes.
This suggests again that interactions between neighboring oxygen atoms might be 
the cause for an increased
energy gap, because disk-like Si-QD have by construction two planar 
surfaces (see Fig.\ \ref{Fig1}) where oxygen neighbors are closer and can interact more 
easily
than in the other two geometries.
Notice that point (5) is complementary to point (3), above. 
Point (3) showed that 
reducing the size of dots corresponds to reducing 
oxygen neighbor interactions due to
curvature effects,
thus reducing the energy gaps compared to that of equivalent hydrogenated Si-QD. 
Point (5) arrives at the same conclusion using equivalent dot sizes but with variable shapes.
These qualitative effects on energy gap variations
due to neighboring oxygen atoms based on these
simplified uniformly passivated surface models
are quantified next using the more realistic 
second set of models that contains the three types of Si-suboxides, as in XPS. 

The second set of models was designed here in order to study effects of neighboring Si-suboxides
at planar surfaces, 
using identical Si-suboxide atomic configurations for various geometries.
By construction, the Si-suboxides at the center of the oxydized surfaces of the
Si(100)-NC  model in Fig.\ \ref{Fig2}(b) 
can interact with each others as if they were located in the Si-QW of 
Fig.\ \ref{Fig2}(c),
while Si-suboxides at the periphery of that planar surface
can only interact with Si-suboxides located towards the center of the plane as in
the quantum dot of Fig.\ \ref{Fig2}(a).
In other words, the model in Fig.\ \ref{Fig2}(b) is an hybrid between those of
Fig.\ \ref{Fig2}(a) and (c).
Therefore, if neighboring Si-suboxides do affect the energy gap, as stated using the 
first set of models, then it is expected that the difference between the energy gaps of 
oxygenated and hydrogenated surfaces of the planar Si(100)-NC
model should be intermediary 
to the equivalent 
differences of energy gaps from the Si-QD and from the Si-QW. 
It indeed is the case as given by
Table \ref{Table1}:
The HOMO-LUMO gap of the Si-QD with Si-suboxides is \emph{smaller} than its 
hydrogenated 
counterpart, giving a difference of -0.21 eV.
The band gap of Si-QW with Si-suboxides is \emph{larger} than its hydrogenated  
counterpart, giving a difference of +0.22 eV, also in
agreement with previous calculations using plane waves.\cite{Carrier}
Finally, the HOMO-LUMO gaps of the planar Si(100)-NC with  or without 
Si-suboxides are relatively 
similar, a mere difference of +0.04 eV,  and a value clearly
in between that of the Si-QD and the Si-QW energy gaps.
Notice that one possible explanation for smaller gap variations in the Si(100)-NC of 
Fig.\ \ref{Fig2}(b)
than in the Si-QD of Fig.\ \ref{Fig2}(a) could 
be attributed to a size effect, where a much larger
quantum dot would give smaller decrease of energy gaps due to oxygen than a smaller dot.
However, it is found here that the HOMO-LUMO gap of the model in Fig.\ \ref{Fig2}(b)
does not show a decrease of the 
energy gap compared to its hydrogenated counterpart,
as is usually the case in oxygenated Si-QD,
but instead a slight \emph{increase}.
Furthermore, in the superlattice [Fig.\ \ref{Fig2}(c)],
which is basically the infinite extension of Fig.\ \ref{Fig2}(b),
the energy gap significantly increases due to oxygen.
It thus means that there exist an additional mechanism not related
to the size of the nanocrystal that does increase the energy gap 
in the presence of Si-suboxides, at least at planar surfaces.
The fact that the energy gap increases due to oxygen was
interpreted using the first set of models above
by an
enhanced interaction 
between neighboring 
Si-suboxides at the planar surfaces of the disk-like dot, 
that further delocalizes electronic states and
thus \emph{increases} the energy gap.
This interpretation is further established next by plotting the charge density along the plane of the
Si-suboxides in the Si(100)-NC model of Fig.\ \ref{Fig2}(b).
%FIGURE_4
\begin{figure}[tb]
   \centering
   \includegraphics[width=2.5in]{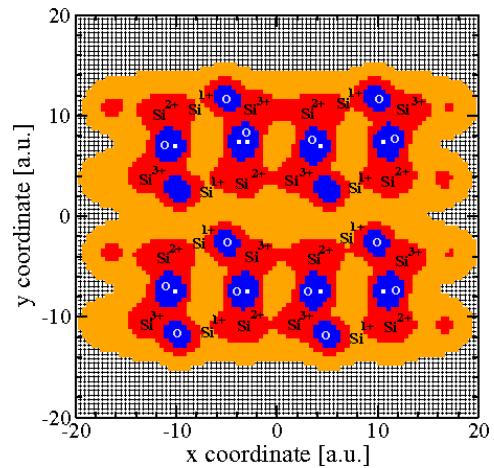} 
   \caption{(Color online) 
Total charge density for the oxygenated Si(100) nanocrystal shown in 
Fig.\ \ref{Fig2}(b), along the plane defined by the Si atoms at the surface.
The grid points where the charge density $\rho$ (in units of electron charge)
is less than 0.001 corresponds to the background mesh (symbol +).
Light (yellow) area: 0.001 $<$ $\rho$ $<$ 0.02.
Darker (red) area: 0.02 $<$ $\rho$ $<$ 0.1.
Darkest (blue) area: 0.1 $<$ $\rho$ $<$ 1.0.
   }
   \label{Fig4}
\end{figure}

It is clear from this study, and from previous publications,
that energy gaps in spherical Si-QD as in Fig.\ \ref{Fig2}(a)
decrease due to an increased localization of the electronic states caused by oxygen atoms.
Any oxygen atom tends to accumulate charge because of its large electronegativity,
equal to 3.44.
Now, the effect of (de)localization due to Si-suboxides at
the planar surface can be established by
depicting the charge density
along the plane formed 
by the Si-suboxides in the model of Fig.\ \ref{Fig2}(b), as shown in Fig.\ \ref{Fig4}.
The first observation from the Figure is that oxygen indeed tend to accumulate charge 
(i.e., the darkest blue regions are on the
oxygen sites in Fig.\ \ref{Fig4} and are the most intense charge densities), as expected.
Most importantly, in addition to these peaks of 
densities located onto the oxygen sites,
the charge distribution surrounding the oxygen charge density peaks is also significant and remains
relatively evenly distributed \emph{around} those peaks, especially at the 
Si$^{3+}$ and the Si$^{2+}$  (the dark red regions in Fig.\ \ref{Fig4}).
It thus suggests that multiplying interactions among Si-suboxides, i.e., 
when there are sufficiently enough Si-suboxides and when they are close enough and are
bridging each others,
then Si-suboxides act as collective entities and
tend to delocalize the electronic
states, from the oxygen sites where the charge remains largely concentrated towards 
the silicon suboxide atoms where the charge becomes relatively evenly distributed.
This slight delocalization of the electronic states then induces an 
increase of the band gap compared to an equivalent 
hydrogenated surface.
Both Si$^{3+}$ and the Si$^{2+}$ seem to be instrumental to the generation of
extended states along that nanosurface.
Inspection of the charge distribution (darker red areas) in the Figure shows that
this distribution is more intense 
at the Si$^{3+}$  and Si$^{2+}$ sites then onto the Si$^{1+}$.
It can be attributed to the fact that, on average (since SiO$_2$ is amorphous in real devices),
Si$^{3+}$ suboxides have more oxygens
bridging other 
silicons than the Si$^{2+}$, and 
furthermore than any of the 
Si$^{1+}$ suboxides.

\section{Conclusions}

This result concerning the relation between energy gap 
and planar surface of few nanometers
in diameter has interesting consequences for measurements of energy gaps in
Si-NC, when planar surfaces are observed at the nano-scale.
For instance, recent photographs of
high resolution transmision electron spectroscopy (HRTEM)
show 
that such nanometer-long planar surfaces can form.\cite{Thogersen}
This first-principles calculation suggests that
planar surfaces in Si-NC may contribute \emph{at least}
0.2 eV to the overall energy gap values in the PL. 
Such blueshift due to interface Si-suboxides may be measurable by comparing PL between 
as-grown samples
and samples annealed at
temperatures at which Si-suboxides are reported to vanish 
(above 950$^{\circ}$C),\cite{Thogersen} assuming no significant alterations to the bulks of the
nanocrystals during annealing.

In summary, 
several models of oxygenated and hydrogenated surfaces of Si confined 
quantum wells and 
nanocrystals of variable shapes have been
constructed in order to determine interface curvature effects on energy 
gaps due to 
Si-suboxides. 
Si-suboxides in quantum dots of spherical shapes are shown to reduce
band gaps compared to hydrogenated dots, consistent with previous published results in 
the literature.
However, this trend is shown to be reversed when planar surfaces are formed, such as in 
Si-NC inside
SiO$_2$ thin films, or in Si quantum wells.
At planar surfaces of few nanometers in diameter collective
interactions among Si-suboxides are enhanced, especially among the
Si$^{3+}$ and Si$^{2+}$, which induces a delocalization of the
electronic states and thus increases the band gap compared to 
equivalent hydrogenated surfaces.

Hybrid density functionals are to be considered in the future for determining 
blueshifts closer to experimental values.
Models of Si-NC with entire surfaces passivated by the three Si-suboxides, as well as 
$p$-$n$ dopants, are under development by the author.

\textbf{Acknowledgements} 
{This work is supported by NSF/DMR-0551195 through the PARSEC project.
The author wants to thank Prof.\ Yousef Saad for providing 
resources and discussions, 
Shuxia Zhang from the Minnesota Supercomputing Institute for her 
assistance with optimizing the PARSEC code 
and the two referees for their fruitful comments.
}

\end{document}